\documentclass[letters,fleqn,usenatbib]{mnras}

\usepackage{newtxtext,newtxmath}

\usepackage[T1]{fontenc}
\usepackage{ae,aecompl}

\usepackage{graphicx}	
\usepackage{amsmath}	
\usepackage{amssymb}	
\usepackage[caption=false]{subfig}
\usepackage{multirow}

\newcommand{\degrees}[1]{${#1}^{\circ}$}
\def\code#1{\texttt{#1}}
\def\Rout{\ensuremath{R_\text{out}}}
\def\Rpeak{\ensuremath{R_\text{peak}}}
\def\Rwall{\ensuremath{R_\text{wall}}}
\def\Rmm{\ensuremath{R_\text{peak,mm}}}    
\def\Rnir{\ensuremath{R_\text{wall,NIR}}}  
\def\Mjup{\ensuremath{\text{M}_\text{Jup}}}

\title[Likely planet-induced gap in T Cha's disk]{A likely planet-induced gap in the disk around T~Cha}
\author[N. P. Hendler et al.]{
Nathanial P. Hendler,$^{1}$\thanks{E-mail: equant@lpl.arizona.edu}
Paola Pinilla,$^{2}$
Ilaria Pascucci,$^{1,3}$
Adriana Pohl,$^{4,5}$
Gijs Mulders,$^{1,3}$
\newauthor
Thomas Henning,$^{4}$
Ruobing Dong,$^{2}$
Cathie Clarke,$^{6}$
James Owen,$^{7}$
David Hollenbach$^{8}$
\\
$^{1}$Lunar and Planetary Laboratory, The University of Arizona, Tucson, AZ 85721, USA\\
$^{2}$Department of Astronomy/Steward Observatory, The University of Arizona, 933 North Cherry Avenue, Tucson, AZ 85721, USA\\
$^{3}$Earths in Other Solar Systems Team, NASA Nexus for Exoplanet System Science.\\
$^{4}$Max Planck Institute for Astronomy, Königstuhl 17, D-69117 Heidelberg, Germany\\
$^{5}$Heidelberg University, Institute of Theoretical Astrophysics, Albert-Ueberle-Str. 2, D-69120 Heidelberg, Germany\\
$^{6}$Institute of Astronomy, University of Cambridge, Madingley Road, Cambridge, CB3 0HA, United Kingdom\\
$^{7}$Institute for Advanced Study, Einstein Drive, Princeton, NJ 08540, USA\\
$^{8}$SETI Institute, Mountain View, CA 94043, USA
}

\pubyear{2017}

\begin{document}
\label{firstpage}
\pagerange{\pageref{firstpage}--\pageref{lastpage}}
\maketitle

\begin{abstract}
We present high resolution ($0.11\arcsec\times0.06\arcsec$) 3\,mm ALMA
observations of the highly inclined transition disk around the star T\,Cha.
Our continuum image reveals multiple dust structures: an inner disk, a
spatially resolved dust gap, and an outer ring.  When fitting
sky-brightness models to the real component of the 3\,mm visibilities, we
infer that the inner emission is compact ($\le1$\,au in radius), the gap
width is between 18-28\,au, and the emission from the outer ring peaks at
$\sim36$\,au. We compare our ALMA image with previously published 1.6$\mu$m
VLT/SPHERE imagery. This comparison reveals that the location of the outer
ring is wavelength dependent. More specifically, the  peak emission of the
3\,mm ring  is at a larger radial distance than that of the 1.6\,\micron{}
ring, suggesting that millimeter-sized grains in the outer disk are located
further away from the central star than micron-sized grains.  We discuss
different scenarios to explain our findings, including dead zones,
star-driven photoevaporation, and planet-disk interactions. We find that
the most likely origin of the dust gap is from an embedded planet, and
estimate --- for a single planet scenario --- that T\,Cha's gap is carved
by a $1.2\Mjup$ planet.
\end{abstract}




\section{Introduction}

Transition disks are a sub-set of disks that display a significantly reduced
near-infrared emission but large mid- to far-infrared emission in their
spectral energy distribution \citep[SED,][]{Strom1989}. This
SED-identified characteristic is associated with a depletion of warm dust particles in
the inner disk, which is why transition disks are thought to be crucial to
understand how planet-forming material is cleared out.

Recently, many transition disks have been spatially resolved at different
wavelengths reveling a variety of structures: gaps or cavities\footnote{We use
the term ``gap'' to refer to an empty or depleted annular region separating an
inner and an outer disk  while we use ``cavity'' for an empty or depleted
region that extends from the central star out to an outer disk.}
\citep[e.g.][]{Perez2014,2017ApJ...836..201D}, spiral arms
\citep[e.g.][]{Muto2012}, shadows \citep[some of them variable with time,
e.g.][]{Pinilla2015}, and lopsided asymmetries
\citep[e.g.][]{vanderMarel2013,Casassus2013}.  These observations
highlight that the SED-classified transition disks are a heterogeneous group of
objects \citep[e.g.][]{2014prpl.conf..497E}, only a subset of which may be
truly caught in the act of dispersing \citep[e.g.][]{2017RSOS....470114E}.
For instance, disks shaped by planet-disk interaction are not necessarily dispersing
as planet formation may occur early in disk evolution and be a long lived
process.

The combination of near-infrared (tracing micron-sized particles) and millimeter imagery (mm/cm-sized grains) provides important insights into the origin of structures
in transition disks. In particular, cavities and gaps resulting from pressure
bumps in the gas are most pronounced at millimeter wavelengths as
mm/cm-sized grains concentrate in pressure maxima, while micron-sized dust follows the gas \citep[e.g.][]{2008A&A...480..859B}. In the case of
planet-induced pressure bumps, the location of the peak emission at
near-infrared and mm wavelengths can be also used to estimate the planet mass
\citep[e.g.][]{2013A&A...560A.111D, Garufi2013}.

In this letter, we present ALMA 3\,mm high-angular resolution
($0.11\arcsec\times0.06\arcsec$) observations of the transition disk around
T-Chamaeleontis (T\,Cha). We also compare our data with the similarly
high-resolution ($\sim 0.04\arcsec$) 1.6\,\micron{} SPHERE/VLT imagery recently
published by \cite{2017A&A...605A..34P}.  This comparison makes T\,Cha one of
two systems, along with TW\,Hya \citep{2017ApJ...837..132V}, for which millimeter and near-infrared
emission can be investigated at similar spatial scales (e.g. 2.5-4.3\,au with SPHERE and 1.6-10\,au with ALMA).

T\,Cha is a T-Tauri star \citep[spectral type G8,][]{Alcala1993} located in the
$\epsilon$-Cha association (at $107\pm3$\,pc, \citealt{GaiaCollaboration2016})
with an estimated age between 2-10\,Myr \citep{Fernandez2008, Ortega2009}.  The
presence of a gap in this disk was first inferred via SED modeling
\citep{Brown2007} and then later by the analysis of NIR interferometric data
\citep{2011A&A...528L...6O,2013A&A...552A...4O}.  High-resolution mid-infrared
spectroscopy found evidence for a disk wind beyond the dust gap whose
properties are compatible with a slow star-driven photoevaporative wind
\citep{2009ApJ...702..724P,2012ApJ...747..142S}.  ALMA Cycle 0 observations at
0.85\,mm could not resolve the gap but identified two emission bumps separated
by 40\,au, suggesting a cavity of 20\,au in size \citep{Huelamo2015}.
Interestingly, a candidate exoplanet has been reported inside the cavity
\citep{Huelamo2011}, although its existence is debated \citep{Sallum2015}. The
recent SPHERE/VLT scattered polarized light images \citep{2017A&A...605A..34P}
resolve the outer ring-like emission. By fitting  the observations with
radiative transfer models, the inner edge of the outer ring is found at
$\sim30\,$au.  \cite{2017A&A...605A..34P} also give upper limits for potential
embedded planets. Using hot-start models, planets more massive than
$\sim8.5\,\Mjup$ are ruled out at a distance from 0.1\arcsec to 0.3\arcsec
(10.7 to 32.1\,au) from the central star.  At larger separations, the limit is
$\sim2.0\,\Mjup$.

Here, we present our ALMA 3\,mm
observations (Sect.~\ref{sect:obs}), then our visibility fitting approach
(Sect.~\ref{sect:analysis}), and summarize our results, including the comparison with the
SPHERE/VLT images (Sect.~\ref{sect:results}).
We discuss our findings and address different origins for
shaping the wavelength dependent size of gaps and cavities (Sect.~\ref{sect:discussion}). 
We conclude that the most likely explanation for T\,Cha's gap is from planet-disk
interaction.

\section{Observation and Data Reduction}\label{sect:obs}

Our ALMA Cycle~3 observations (Project ID: 2015.1.00979.S, PI I. Pascucci) were
carried out on October 27 and 29, 2015 with 48 12\,m antennas.  Of the four
Band~3 spectral windows, three were utilized for the continuum emission and one was
centered around the hydrogen recombination line H(41)$\alpha$ at 92.03\,GHz,
which might trace a disk wind
\citep{2012ApJ...751L..42P}. The on-source time ($\sim$2\,h) was set to
achieve a sensitivity of 9$\mu$Jy/beam in the aggregate continuum bandwidth of 6.6\,GHz.

\begin{figure*}
\centering
\subfloat[]{%
    \includegraphics[width=\textwidth]{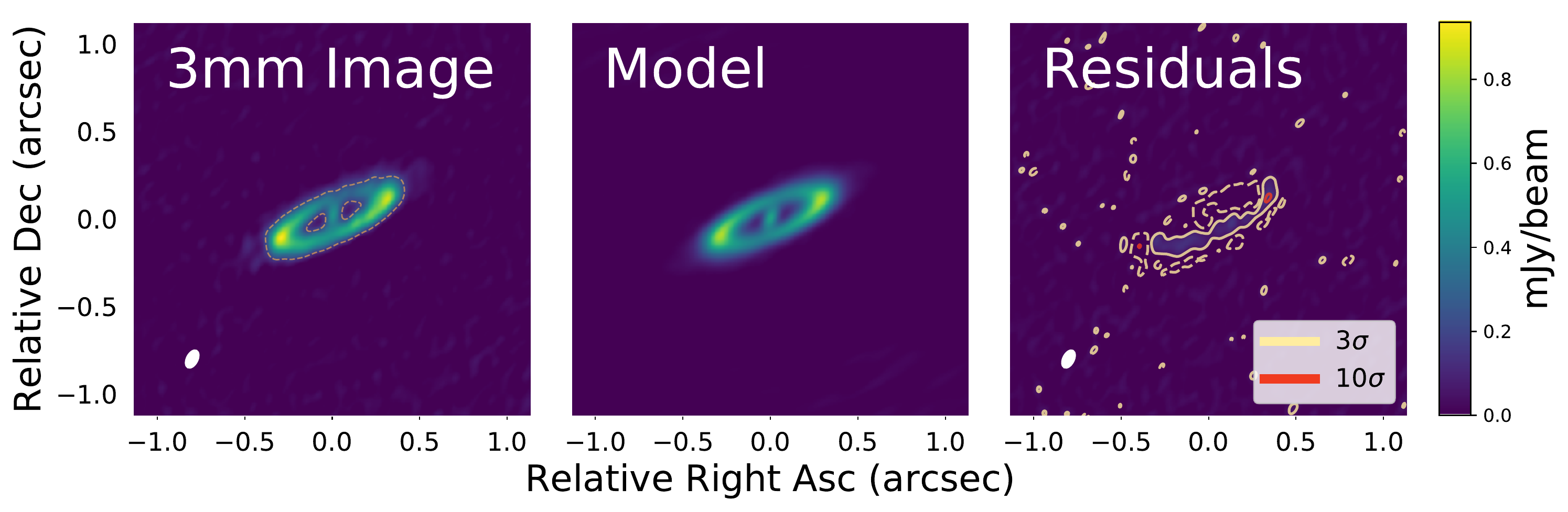}%
}
    \caption{
        Left: Cleaned 3\,mm ALMA image of T\,Cha (Project ID: 2015.1.00979.S).
        Beam is shown as ellipse in lower left and $1\sigma$ level is 1.6mJy.  Countour shows $5\sigma$ region integrated for flux density measurement reported in Section~\ref{sect:obs}.
        Center: Best-fit model.
        Right: The residuals of our best-fit model (center panel) subtracted from the unbinned ALMA image (left panel).
    }\label{fig:images}
\end{figure*}

The ALMA data were calibrated using the Common Astronomy Software Applications
\citep[CASA,][]{2007ASPC..376..127M}.  The initial reduction scripts were
provided by the North American ALMA Science Center and included phase,
bandpass, and flux calibration. We ran the scripts using CASA 4.7.1 and
created images in the continuum and in the line using Briggs weighting and
robustness parameter equal to 0.5 (for the line we also applied a 2M$\lambda$
tapering to increase the sensitivity). The continuum emission at an average
frequency of 99.16\,GHz (3.0\,mm) is spatially resolved and detected at high
sensitivity, see Figure~\ref{fig:images} where the image rms is
9$\mu$Jy/beam and the beam is $0.11'' \times 0.06 ''$ with a position angle of
-18.11$^{\rm o}$.  The H recombination line is not detected even when rebinning in
spectral resolution to tens of km/s, the expected line width. Applying
self-calibration did not significantly improve the continuum image while
resulting in lower spatial resolution and did not lend a line detection.
Hence, we will focus this letter on the properties of the continuum 3\,mm
emission using the non self-calibrated data.

On the image shown in Figure~\ref{fig:images} we draw a 5$\sigma$
contour and measure a total flux density of 16.5$\pm$1.6\,mJy (the quoted
uncertainty includes an absolute flux calibration uncertainty of 10\%). The peak S/N from our cleaned image is 110.  Our
value is $\sim$2.5 higher than that found by ATCA at 3.2\,mm
\citep{2012MNRAS.425.3137U}, which, in combination with the flux reported by
\cite{Huelamo2015} (198$\pm$4\,mJy at 0.85\,mm), results in a integrated
millimeter spectral index of $\sim1.97\pm0.08$. This value is close to the relationship between
spectral index and cavity size introduced by \cite{2014A&A...564A..51P} for
transition disks 
and could be an indication of mm-sized grains
trapped at the outer edge of a planet-induced gap. 

\begin{figure}
\centering
    \includegraphics[width=\columnwidth]{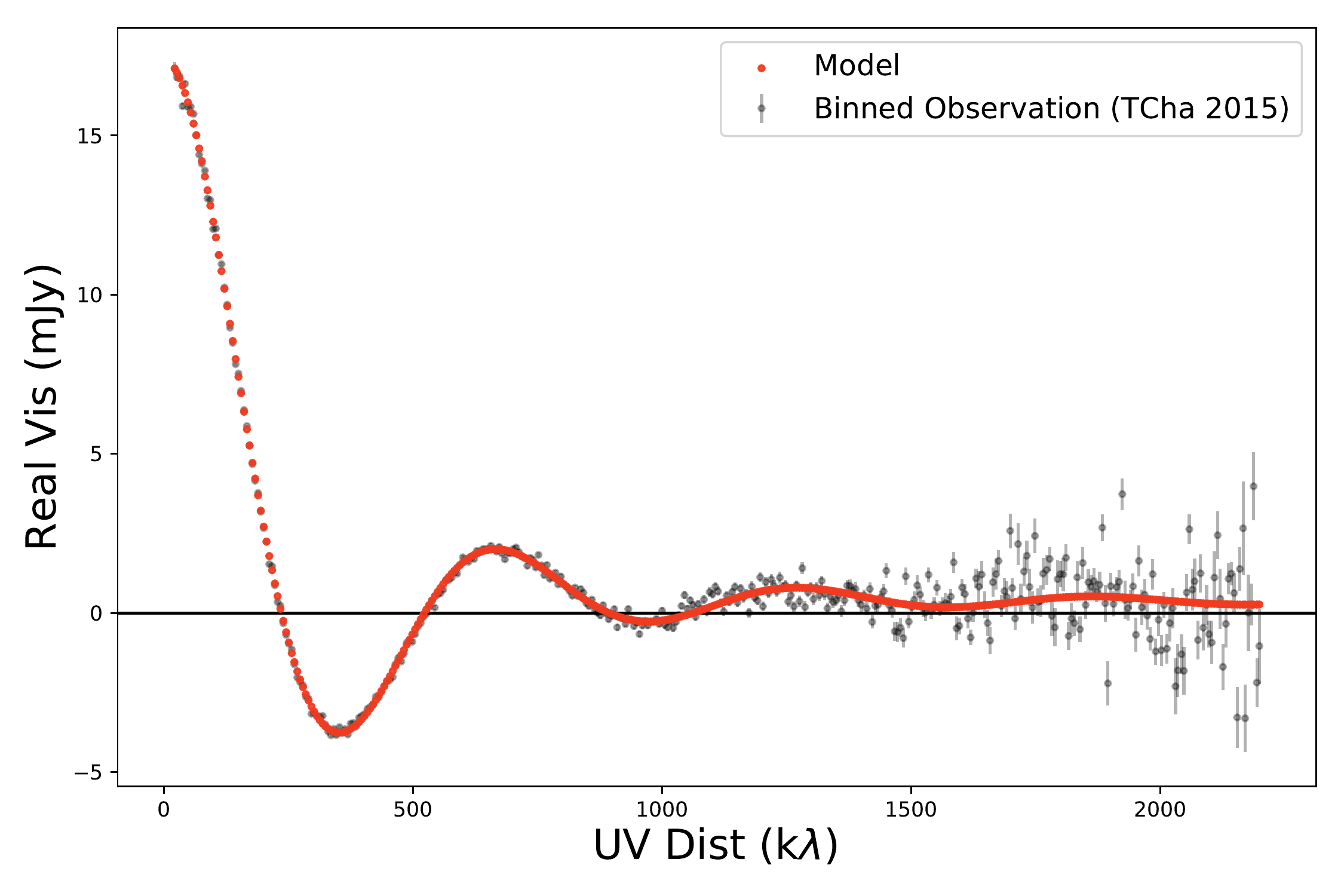}%
    \caption{
    Deprojected real visibilities from Band 3
    observation.  Data are shown in gray with error bars.  Model is
    overlaid in red.
    }\label{fig:visibilities}
\end{figure}

\section{Analysis}\label{sect:analysis}

The T\,Cha 3\,mm continuum map (hereafter: 3\,mm\, image,
Figure~\ref{fig:images}) reveals a disk with three distinct
components: an inner emission, a spatially resolved gap, and an outer ring.  In
order to quantify the location and size of these structures we use two methods
-- fitting model radial profiles to the
visibilities and direct measurements from the 3\,mm image.  We then compare our results with previously published findings at shorter
wavelengths in order to examine the wavelength dependence of these features.

\begin{table}
\begin{center}
\begin{tabular}{llccc}
\hline \noalign {\smallskip}
    Symbol    & Parameter                            & Range    & Best Fit     & Unit \\ \hline
    \multicolumn{5}{c}{Nuker Profile -- Outer Disk} \\
    $\Rpeak$    & transition radius & [10--60] & $37.09  \pm 0.07$  & (au) \\
    $\gamma$ & inner disk index  & [-6--1]  & $-3.10  \pm 0.06$  & \\
    $\alpha$ & transition index  & [10--85] & $53.73  \pm 12.81$ & \\
    $\beta$  & outer disk index  & [5--8]   & $6.49   \pm 0.04$  & \\
    \hline
  \multicolumn{5}{c}{Gaussian -- Inner Disk} \\  
    A/B              & amplitude ratio & [30--500] & $336.0  \pm 109.1$ & \\
    $R_\text{width}$ & gaussian width  & [0--4]    & $1.01   \pm 0.28$  & (au) \\
    \hline
\end{tabular}
    \caption{\em{Model parameters, parameter space (Range), and best fit modeling results.}}\label{tab:model_parameters}
\end{center}
\end{table}

Assuming an axisymmetric disk, we model the sky brightness by fitting
parametric models to the dust continuum emission in the visibility domain.  To
prepare the visibilities for fitting, we center, deproject and bin the data.
The visibilities were centered using the routines \code{uvmodelfit()} and
\code{fixvis()} within CASA.  CASA's \code{disk} model provided the best
centering fit, resulting in a center with $\alpha_{ICRS}=$11h57m13.29s and
$\delta_{ICRS}=$-79d21m31.68s.
The observed ($u$, $v$) points from the three spectral windows are deprojected
and binned, reducing the number of unflagged data points used for fitting from
$7\times10^7$ to $3\times10^4$.  During this process, we minimize the residuals
between the unbinned image and our axisymmetric binned image in order to find a fit for the
inclination and position angle (PA).  We find that the optimal values for the inclination and PA are
\degrees{73} and \degrees{113}, respectively, which are within 2$\sigma$ of the
values and uncertainties reported in the literature
\citep[e.g.][]{2013A&A...552A...4O,2017A&A...605A..34P}.

To model the three disk components, we assume radial intensity profiles that
combine a Nuker profile and a Gaussian.  \cite{2017ApJ...845...44T} showed that
a Nuker profile \citep{1995AJ....110.2622L} reproduces well the visibilities of
a variety of disks,
including
full and transition disks, with few input parameters.
It does this by producing a decreasing monotone function for full disks, or a
symmetric/asymmetric function about a single maximum for rings.  The addition of a Gaussian centered at the stellar location
is necessary to fit the inner emission seen in
Figure~\ref{fig:images}.  The radial profile of our model is given by

\begin{equation} \label{eq:radial_profile}
    I(r) = A~\text{exp}\bigg(-\frac{r^2}{2R_\text{width}^2}\bigg) +  B\bigg(\frac{r}{\Rpeak}\bigg)^{-\gamma}\bigg[1 + \bigg(\frac{r}{\Rpeak}\bigg)^{\alpha}\bigg]^{(\gamma-\beta)/\alpha}
\end{equation}

where the first term in the equation is the central Gaussian and the second term is the Nuker profile.
In the equation, $r$ is the radial distance and the remaining symbols are defined in
Table~\ref{tab:model_parameters}. When Fourier transformed this symmetric
brightness profile can be expressed with the zeroth-order Bessel function of
the first kind $J_0$ \citep{2007NewAR..51..576B}.
We then fit the model visibilities to the
binned real part of the observed visibilities using the $emcee$
\citep{2013PASP..125..306F} implementation of the Markov chain Monte Carlo
method.  For our radial grid, we use $r\in[0-500]$ au with steps of 0.025\,au.
Our parameter space is sampled with 1000 walkers having 500 steps each.  The
parameter space explored, and the results of our fitting are shown in
Table~\ref{tab:model_parameters}.  Posterior probability distributions of the
sampled parameter space are available online.

Our best-fit model is imaged using the same ($u$, $v$) coordinates as our ALMA
observations.
Our model visibilities are shown in Figure~\ref{fig:visibilities},
while Figure~\ref{fig:images} shows the model image 
and the residuals between our unbinned observations and our best-fit model, i.e.
$\text{data}-\text{model}$.
Note that the residuals are at most 10$\sigma$,
and below 1$\sigma$
when we subtract the model from the binned data.
Our residuals reveal
a positive residual in the south spanning from east-to-west, and negative
residuals in the south and north . Because positive and negative residuals are
aligned with the disk major axis, we argue that this may result from
optical-depth effects.

\section{Results}\label{sect:results}

\begin{figure*}
\centering
    \subfloat[]{%
    \includegraphics[width=0.45\textwidth]{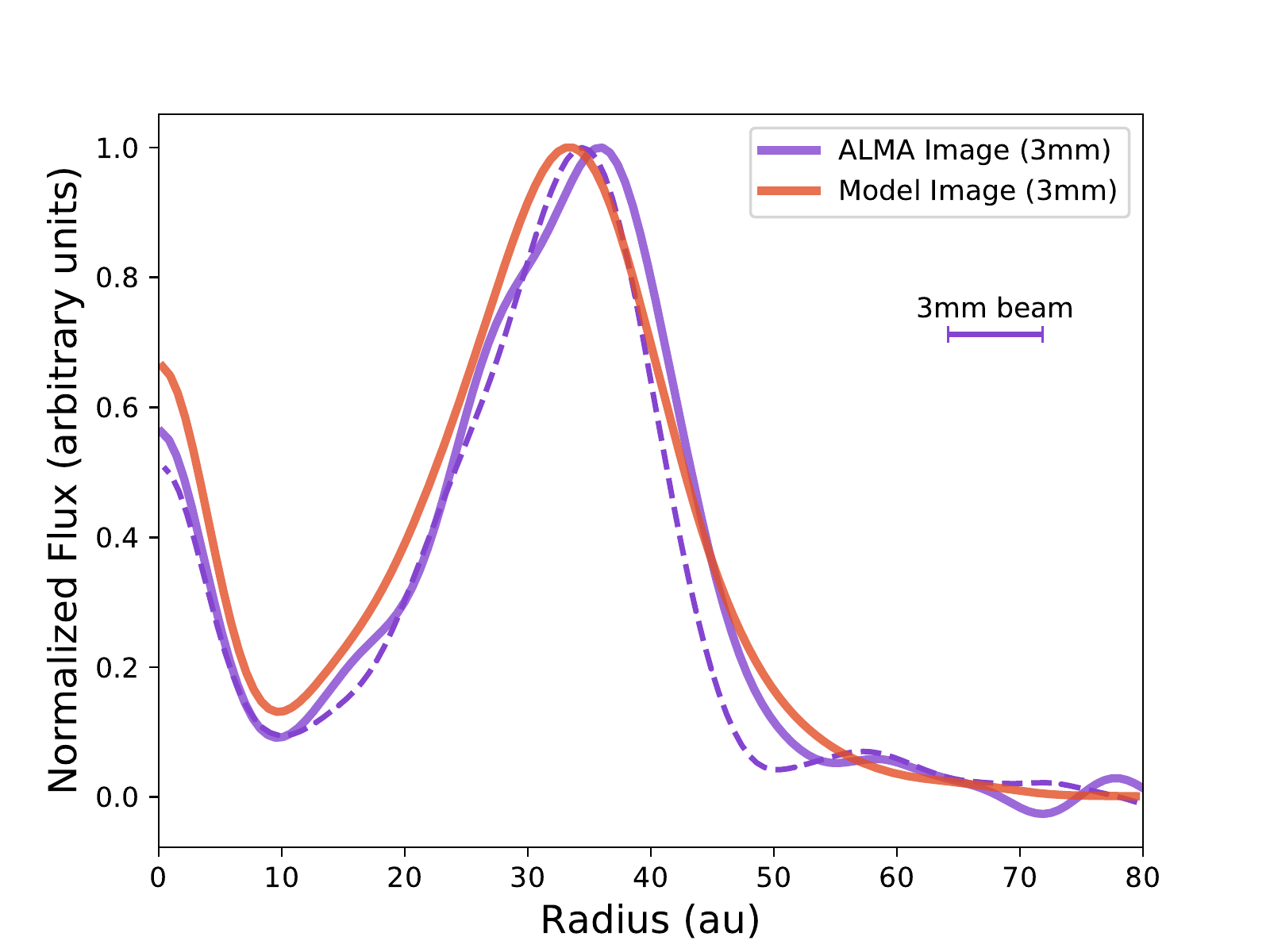}%
    \label{fig:image_radial_profiles}%
}\qquad
    \subfloat[]{%
    \includegraphics[width=0.45\textwidth]{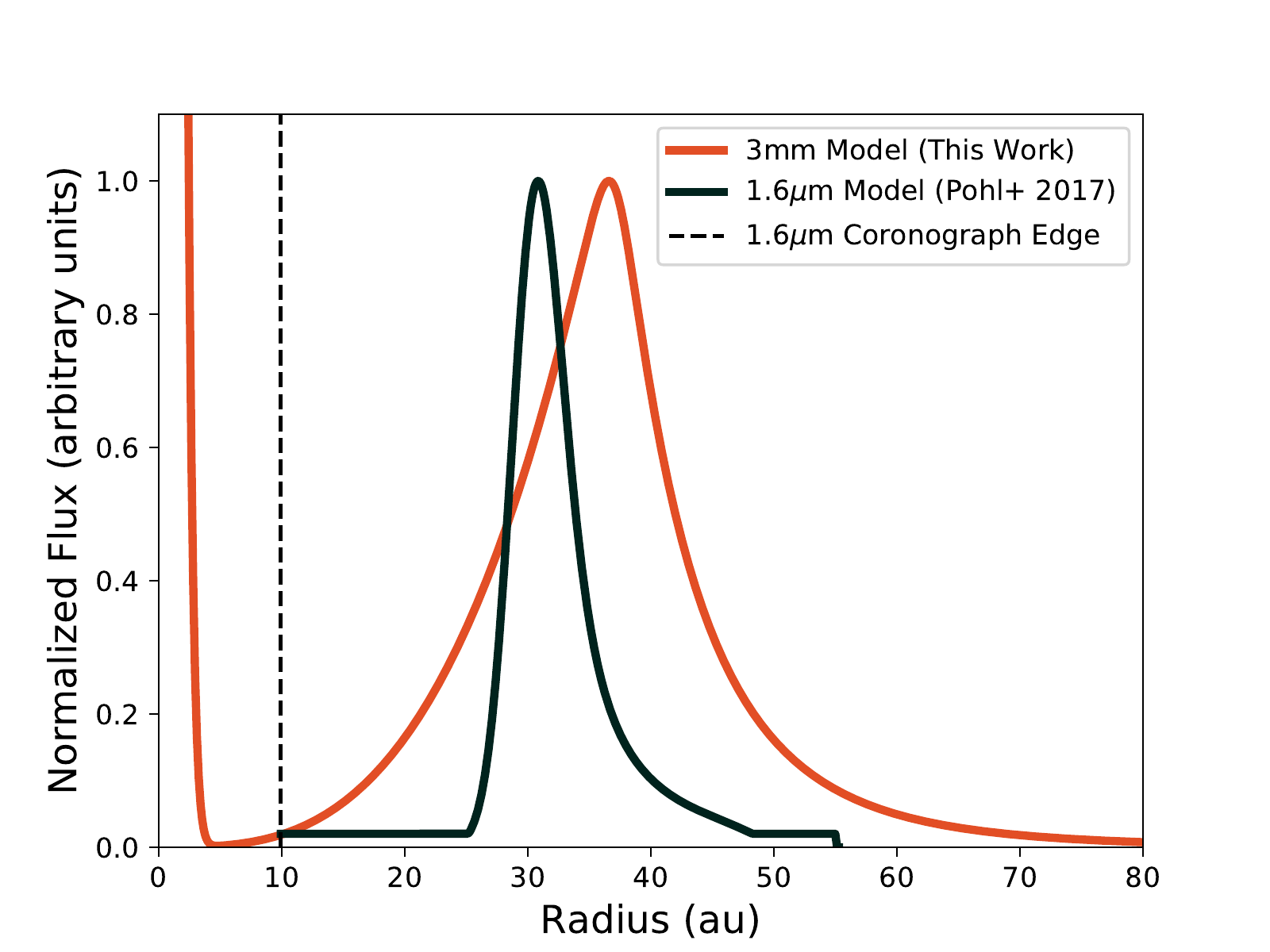}%
    \label{fig:model_radial_profiles}%
}
    \caption{
        (a) Comparison of radial intensities profiles (normalized to the intensity of the ring peak)
        from our ALMA image and best-fit model
        image (using same $u$,$v$ spacing).  For the ALMA image (purple), the solid
        line is taken radially along T\,Cha's PA towards the NE, while the dashed
        lines is taken along the same line extending to the SW.  Because our
        model is axisymmetric, we show only one profile (red).
        The ALMA/3mm beam width along T\,Cha's PA is $0.072\arcsec$.
        (b) Comparison of ring location at different wavelengths.  Radial
        intensities taken from best-fit models.
    }
    \label{fig:profiles}
\end{figure*}

Radial intensity profiles along the disk's PA of \degrees{113} are shown in
Figure~\ref{fig:image_radial_profiles} for our ALMA 3\,mm image. We also
compare in Figure~\ref{fig:model_radial_profiles} the best-fit model profiles
from our visibility analysis and the best model fit to the  SPHERE/VLT
$1.6\,\micron$ image from \cite{2017A&A...605A..34P}. From these profiles we
extract values for the location of the ring's peak emission ($R_\text{peak}$).
In addition to \Rpeak, we report \Rwall{} defined as in
\cite{2013A&A...560A.111D}, i.e. the radial location
where the intensity is half of the difference between the gap
minimum intensity and the outer ring peak intensity. The radial position of \Rpeak{}
resulting from our 3mm ALMA observations agree well (within 1\,au) with our
best-fit model (see Table~\ref{tab:model_results}).  Comparing our visibility
modeling with the NIR model provided by \cite{2017A&A...605A..34P}, we find
that the 3mm \Rpeak{} is at a greater radial distance than the 1.6$\mu$m
\Rwall{} by $8.8$\,au.

The central emission is unresolved by
our beam of $\sim$10\,au diameter and our visibility modeling suggests that it
is confined within a radius of only $\sim 1$\,au.  The flux density measured
over the ALMA beam is $0.55\pm0.06$\,mJy. By extrapolating the hot (1400K) NIR
SED-component \citep{2013A&A...552A...4O} out to 3\,mm, we find that only $\sim$2\%
of the measured flux density could come from close-in micron-sized grains. In
addition, from the long-wavelength portion of the SED
\citep{2014ApJ...795....1P} we calculate that no more than $\sim30\%$ of the
central emission arises from free-free emission. As such, most of the detected
central emission is thermal dust emission, likely from millimeter-sized grains.
By detecting the central 3\,mm emission {\it and} spatially resolving the outer
ring, our ALMA image demonstrates that the disk of T\,Cha has a gap at
millimeter wavelengths and not a cavity.

Using the definition of \Rwall{} for the locations of the inner and outer edge
of the gap, we estimate from our image a gap size of $\sim 17.7$\,au.  This
should be considered a lower limit as we do not spatially resolve the central
mm emission.  The location of the gap's outer edge found from our observation
and modeling are 22.9 and 27.8\,au respectively, with the latter being very
similar to the 28.3\,au reported by \cite{2017A&A...605A..34P} from modeling
the SPHERE images.  By taking the model \Rwall{} to be the location of
the gap outer edge, we find an upper limit on the gap width of $< 27.8$\,au
(any larger and the gap would extend to zero au and become a cavity).

\begin{table}
\begin{center}
\begin{tabular}{llc}
\hline \noalign {\smallskip}
    $\lambda$ &
    Source &
    Location \\
\hline
    \multirow{3}{*}{3mm}         & Image \Rpeak & 36.1$\pm$0.8\,au \\
                                 & Model \Rpeak & 37.1\,au \\
                                 & Image \Rwall & 22.9\,au \\
                                 & Model \Rwall & 27.8\,au \\
    \hline
    \multirow{2}{*}{1.6$\mu$m}   & Model \Rpeak & 30.8\,au \\
                                 & Model \Rwall & 28.3\,au \\
    \hline
\end{tabular}
    \caption{\em{Radial locations of the outer-ring emission found from images and best-fit models.}}\label{tab:model_results}
\end{center}
\end{table}

\section{Discussion}\label{sect:discussion}

Our continuum ALMA observations reveal a disk around T\,Cha with three main
dust components: an inner emission, gap, and an outer ring. By modeling the real part of the visibilities with a Nuker profile and a Gaussian, we can constrain the physical properties of these three components.

Our main result is the detection of an unresolved inner disk ($\le\,1$\,au radius, based on visibility modeling), where most of the 3\,mm flux density arises from thermal dust emission, likely from mm-sized grains (see Sect.~4). This, combined with the already known outer emission, implies that the disk of T~Cha has a gap in the population of millimeter-sized grains.
Previous work has also inferred a very compact inner disk in the NIR \citep[$\Rout
\le 0.17$\,au,][]{2011A&A...528L...6O,2013A&A...552A...4O}. 
Hence, whatever is the physical origin of the gap, it has to allow a detectable inner disk at multiple wavelengths.


The detection of a gap, and not of a cavity, in combination with other
properties inferred from our analysis, enables us to exclude several mechanisms
for the origin of the structures in the disk of T\,Cha.  For instance, particle trapping at the outer edge of a dead zone can create structures as observed in transition disks \citep[e.g.][]{2015A&A...574A..68F,2016A&A...590A..17R}. Inside a dead zone, where turbulent velocities of particles are low, grain growth is efficient and small grains are depleted.  As a result, we expect to observe a NIR and a mm-cavity of the same size.  When an MHD disk wind
acts together with a dead zone, smaller cavities will be present at  shorter
wavelengths.  However, this latter scenario does not preserve a long-lived inner
disk because the MHD wind removes efficiently the inner material, leaving a
cavity, not a gap \citep{2016A&A...596A..81P}. As a
consequence, a dead zone, or a dead zone \emph{and} MHD disk wind, is not a likely origin
for the observed \emph{gap} in the disk of T\,Cha.


A gap may originate from central star-driven photoevaporation \citep[second
disk evolutionary stage in Figure~6 of][]{2014prpl.conf..475A}.  While there is
evidence in T~Cha of a disk wind that can be photoevaporative in nature
\citep{2009ApJ...702..724P,2012ApJ...747..142S}, photoevaporation predicts only
a specific combination of mass accretion rate and gap size \citep[see Figure~6
in][]{2017RSOS....470114E}.  T\,Cha shows strong photometric variability and UX
Ori-like behavior, inferred from significant changes of prominent emission
lines, such as $H_\alpha$. If this variability is due to accretion episodes,
the mean accretion rate of T\,Cha is $4\times10^{-9}\,M_\odot$/year
\citep{Schisano2009}. With this mass accretion rate and our gap size estimate
$\ge$20\,au, star-driven photoevaporation does not appear to be a plausible
mechanism to open the gap seen in the T~Cha disk.  This argument is valid even
in the presence of variability because the timescale for the hole to grow from
the initial gap opening radius to $\sim$20\,au is longer than the timescale for
draining material by accretion.  If the accretion rate in the outer disk is
variable then there will be episodes when the inner disk refills and there is
accretion onto the star, which will reduce the gap size. 


Both cavities and gaps are expected from planet-disk interactions.  While a
cavity can be opened by a massive planet, \cite{2016A&A...585A..35P}
demonstrated that an inner disk can be maintained at timescales greater than
1\,Myr for planets of mass $\sim1.0\,\Mjup$. More massive planets filter out
all kind of grains (from micron- to mm/cm-sized particles), and as a result the
inner disk is gone after few million years of evolution.

The spatial segregation of small and large particles seen in T\,Cha, as
observed in the NIR and mm-emission respectively, is another expected outcome
of planet-disk interactions, because the small grains are expected to follow
the gas while the large grains move and accumulate in pressure maxima.  Indeed,
radial segregation of small vs. large grains has been observed in different
transition disks following the expected results from planet-disk interaction
models \citep{Garufi2013,2015A&A...584A..16P}.  Using the wavelength-dependent
grain size relationships resulting from the 2D hydrodynamical and dust
evolution models of \cite{2013A&A...560A.111D}, we can constrain the mass of
the potential embedded planet inside the gap.  With the values reported in
Table~\ref{tab:model_results}, we find $\Rnir/\Rmm = 0.8$ which implies a
$1.2\,\Mjup$ planet in the single-planet scenario and with the specific disk
properties assumed in \cite{2013A&A...560A.111D} (model outcomes are sensitive
to disk properties, especially viscosity).
The planet mass is consistent with the upper limits reported in
\cite{2017A&A...605A..34P}, leaving open the possibility that the gap in the
disk of T~Cha is carved by a giant planet.

While a large millimeter gap can be consistent with a single planet,
\cite{2017ApJ...835..146D} show that the width of a single-planet gap observed at
near-infrared wavelengths should be only $\sim 30-40\%$ of the gap radial
location.  Using $\Rnir$ from Table~\ref{tab:model_results} and the coronagraph
mask diameter of 185\,mas gives a minimum NIR gap width  of $\sim18$\,au, which
is $\ge$64\% of the gap's radial location. As such it is unlikely that the gap
in the disk of T~Cha is opened only by one planet. Rather multiple embedded
planets, whose individual gaps overlap, contribute to open the large observed
gap.

In summary, we find that the most likely origin for the gap in the disk of
T\,Cha is due to planet-disk interactions.  If the gap is carved by a
single planet, we find that a mass of $1.2$\,\Mjup{} is required.  However,
because of the large gap width, multiple less massive planets with overlapping gaps are more probable.  Future ALMA observations could constrain
the size of a potential gap in the gas surface density \citep[e.g.][]{2016A&A...585A..58V}, which is crucial to
distinguish between single or multiple planets \citep[e.g.][]{2011ApJ...729...47Z}.

\section*{Acknowledgements}

We thank the anonymous referee for useful comments and suggestions.
I.P. and N.H. acknowledge support from an NSF Astronomy \& Astrophysics
Research Grant (ID: 1515392). P.P. acknowledges support by NASA through Hubble
Fellowship grant HST-HF2-51380.001-A awarded by the Space Telescope Science
Institute, which is operated by the Association of Universities for Research in
Astronomy, Inc., for NASA, under contract NAS 5-26555.  This paper makes use of
the following ALMA data: ADS/JAO.ALMA\#2015.1.00979.S. ALMA is a partnership of
ESO (representing its member states), NSF (USA) and NINS (Japan), together with
NRC (Canada), MOST and ASIAA (Taiwan), and KASI (Republic of Korea), in
cooperation with the Republic of Chile. The Joint ALMA Observatory is operated
by ESO, AUI/NRAO and NAOJ. The NRAO is a facility of the NSF operated under
cooperative agreement by Associated Universities, Inc.




\bibliographystyle{mnras}





%
%


\bsp	
\label{lastpage}
\end{document}